\documentclass[twocolumn,floatfix,pra,showpacs,floatfix]{revtex4}
\usepackage{graphicx}
\usepackage{color}
\usepackage{psfrag}
\usepackage{epsfig}
\usepackage{bbm}
\usepackage{bm}
\newcommand{\ket}[1]{| #1 \rangle}

\newcommand{\rb}[1]{\left( #1 \right)}

\newcommand{\ew}[1]{\langle #1 \rangle}
\newcommand{\beq}{\begin{eqnarray}}
\newcommand{\eeq}{\end{eqnarray}}

\newcommand{\svec}{\mbox{\boldmath$\sigma$}}

\newcommand{\sfrac}[2]{\begin{array}{c}\frac{#1}{#2}\end{array}}

\begin{document}
\title{Optically-controlled single-qubit rotations in self-assembled InAs quantum dots}
\author{C.~Emary and L.~J.~Sham}
\affiliation{Department of Physics, University of California San
Diego, La Jolla, CA 92093, U.S.A.}
\date{9th January 2007}
\begin{abstract}
We present a theory of the optical control of the spin of an
electron in an InAs quantum dot.  We show how two Raman-detuned laser
pulses can be used to obtain arbitrary single-qubit rotations via
the excitation of an intermediate trion state.
Our theory takes into account a finite in-plane hole $g$-factor and hole-mixing.
We show that such rotations can be performed to high fidelities with
pulses lasting a few tens of picoseconds.

\end{abstract}
\maketitle
%%%%%%%%%%%%%%%%%%%%%%%%%%%%%%%%%%%%%%%%%%%%%%%%%%%%%%%%%%%%%%%%%%%%%%%%
The spin of an electron in a quantum dot (QD) is currently viewed as one
of the leading contenders in the drive to develop a qubit from which
a practical quantum computer can be constructed \cite{zol05}.
The optical manipulation of such spins is an area of tremendous
contemporary interest and offers a route to the ultra-fast control
of qubits required to implement quantum logic.

In this paper we provide a theory of optically-controlled
single-qubit rotations of a single electron spin in a
self-assembled InAs QD.  Such dots are at the forefront of
experimental effort in this field, and their promise for quantum
computation purposes has been illustrated in a number of recent
experiments.
For example, the spin state of a QD electron has been prepared in a
pure state to a very high-fidelity \cite{ata06}, and the
exchange-interactions between optically-excited coupled dots have
recently been mapped \cite{sch06}.

The theory we describe here shows that by using two pulsed,
Raman-detuned lasers one can obtain arbitrary single-qubit rotations
of the electron spin.  The rotation proceeds through the virtual excitation of an intermediate trion state.
We show that, for realistic parameters, this
can be accomplished with laser pulses lasting just a few tens of picoseconds.
This is significantly faster than the decoherence time of the electron spin in such a dot, which has been measured at a few microseconds \cite{gre06}. Furthermore, since the trion state is only virtually excited, this technique avoids problems arising from spontaneous emission from this state.

This work is an extension of that of Chen {\it et al.} \cite{che04},
who have presented a theory of single-qubit rotations for dots in
which the heavy holes have a zero in-plane $g$-factor.  Subsequent
experiments \cite{xia06} have shown that, although this is the case
in GaAs fluctuation dots \cite{tis02}, the $g$-factor in
self-assembled InAs dots is actually finite.  This
is important because an in-plane magnetic field is vital to the rotation
mechanism. Furthermore, in InAs dots
we expect significant hole-mixing, and this also needs to be taken
into account.
Finally, the analytic technique that we use here is somewhat
different from that employed in Ref.~\cite{che04}, and is capable of
describing the qubit rotations to a much higher fidelity.

\section{Spin-trion qubit system}

%%%%%%%%%%%%%%%%%%%%%%%%%%%%%%%%%%%%%%%%%%%%%%%%%%%%%%%%%%%%%
\begin{figure}[t]
  \begin{center}
  \psfrag{t+}{{\Large$\ket{\tau+}$}}
  \psfrag{t-}{{\Large$\ket{\tau-}$}}
  \psfrag{x+}{{\Large$\ket{x+}$}}
  \psfrag{x-}{{\Large$\ket{x-}$}}
  \psfrag{Eze}{{\Large$E_B^e$}}
  \psfrag{Ezh}{{\Large$E_B^h$}}
  \psfrag{V1}{{\Large$\mathbf{V}_1$}}
  \psfrag{V2}{{\Large$\mathbf{V}_2$}}
  \psfrag{H1}{{\Large$\mathbf{H}_1$}}
  \psfrag{H2}{{\Large$\mathbf{H}_2$}}
  \psfrag{d}{{\Large$\delta$}}
  \epsfig{file=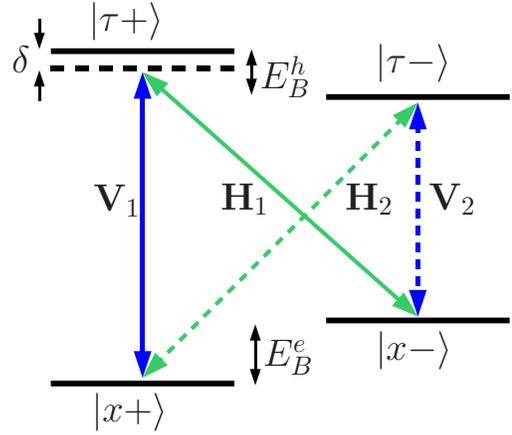, clip=true,width=0.8\linewidth}
  \caption{
    The four-level model of the electron-trion system in the Voigt basis consists of
    two Zeeman-split single-electron ground states $\ket{x \pm}$
    with spins aligned in the $x$ direction,
    and two trion levels $\ket{\tau \pm}$ with
    heavy-hole spins also in the $x$ direction.
    The ground states are split by Zeeman energy $E_B^e$ and the
    trion states by $E_B^h$.
    Arrows indicate allowed optical transitions with $\mathbf{H}$ and $\mathbf{V}$ denoting
    two orthogonal linear polarizations.
    To obtain single-qubit rotations we pump transitions $\mathbf{V}_1$ and $\mathbf{H}_1$
    with a common Raman detuning of $\delta$.
    \label{f1}
 }
  \end{center}
\end{figure}
%%%%%%%%%%%%%%%%%%%%%%%%%%%%%%%%%%%%%%%%%%%%%%%%%%%%%%%%%%%%%%%%%%

We consider a singly-charged self-assembled InAs QD with growth
direction $z$.
The spin of the electron trapped in the QD is our qubit degree of
freedom and we will perform rotations of this spin through the
virtual excitation of an exciton within the dot.
Figure \ref{f1} shows the four-level model that describes the
pertinent features of this system.
We apply a magnetic field in the $x$-direction.  The Zeeman energy
of a QD electron in this field is ${\cal H}_B^e = g_x^e \mu_B B_x
s^e_x \equiv E^e_B s^e_x$, where $g_x^e$ is the electronic
$g$-factor, $\mu_B$ is the Bohr magneton,  $B_x$ is the magnitude of
the field, and  $s^e_x = \pm 1/2$ corresponds to the electron spin.

The heavy-hole component of the trion also splits under this field
in InAs QDs \cite{xia06, ce06}, and can be described with a Zeeman Hamiltonian ${\cal
H}_B^h = - g_x^h \mu_B B_x s^h_x \equiv E^h_B s^h_x$, where
$s^h_x=\pm 1/2$ are the eigenvalues of a pseudo-spin, the components
of which correspond to heavy-hole states aligned in the
$x$-direction, and $g_x^h$ is the hole $g$-factor.
Recent measurements have given the magnitudes of
these $g$-factors as $|g_x^e| = 0.46$ and $|g_x^h|=0.29$ \cite{xia06}.
For concreteness, we follow
Ref.~\cite{ce06} here and assume that both these $g$-factors are
negative, but this is in no way essential.
In our initial treatment we will neglect hole mixing and show later
that it can be incorporated into the analysis through a set of only
very minor modifications.

The four levels of our model are then: the two electron ground
states with spins in the $x$-direction, $\ket{x\pm} \equiv
2^{-1/2}\rb{\ket{\downarrow} \pm \ket{\uparrow}}$, where
$\ket{\downarrow}$ and $\ket{\uparrow}$ represent electron spins in
the $z$ direction; and the two trion levels, $\ket{\tau\pm} \equiv
2^{-1}\rb{\ket{\downarrow\uparrow}-
\ket{\uparrow\downarrow}}\rb{\ket{\Downarrow} \pm \ket{\Uparrow}}$,
where $\ket{\Downarrow}=\ket{\sfrac{3}{2},-\sfrac{3}{2}}$ and
$\ket{\Uparrow}=\ket{\sfrac{3}{2},\sfrac{3}{2}}$ denote heavy-hole
states also aligned in the $z$ direction.

Figure \ref{f1} shows the allowed optical transitions between these
levels.  Due to the splitting of the trion level, these transitions
are linearly polarized.  We have defined the polarization vectors in
terms of $\svec_\pm$ circular polarizations as $\mathbf{V} =
2^{-1/2}(\svec_- + \svec_+)$ and $\mathbf{H} = 2^{-1/2}(\svec_- -
\svec_+)$ \cite{ce06}.

To obtain our qubit rotations, we illuminate the system with two
phase-locked laser pulses propagating in the $z$ direction.  We use
one $\mathbf{V}$-polarized pulse with frequency $\omega_V$ and
time-dependent Rabi frequency $\Omega_V(t)$, and one
$\mathbf{H}$-polarized pulse with frequency $\omega_H$ and Rabi
frequency $\Omega_H(t)$.  With these lasers we pump the two
transitions to the trion level $\ket{\tau_+}$, labelled
$\mathbf{V}_1$ and $\mathbf{H}_1$ in Fig.~\ref{f1},  with a common
Raman detuning of $\delta$.
The model Hamiltonian in the basis $\left\{ \ket{x+},
\ket{x-},\ket{\tau+},\ket{\tau_-} \right\}$ is then
\begin{widetext}
\beq
  {\cal H} =
  \rb{
    \begin{array}{cccc}
      +E^e_B/2 & 0 & \Omega_V^* e^{i \omega_V t+i\alpha} & \Omega_H^* e^{i \omega_H t} \\
      0 & - E^e_B/2 & \Omega_H^* e^{i \omega_H t} & \Omega_V^* e^{i \omega_V t+i\alpha} \\
      \Omega_V e^{-i \omega_V t-i\alpha} & \Omega_H e^{-i \omega_H t} & E_\tau+E^h_B/2 & 0 \\
      \Omega_H e^{-i \omega_H t} & \Omega_V e^{-i \omega_V t-i\alpha} & 0 & E_\tau-E^h_B/2
    \end{array}
  },
\eeq
\end{widetext}
where $E_\tau$ is the trion energy, $\alpha$ is the relative phase
of the two lasers and we have set $\hbar =1$.  The time-dependence
of $\Omega_V$ and $\Omega_H$ is understood.

We set the frequencies of the two lasers as
$  \omega_V = E_\tau - \frac{1}{2}\Sigma_B-\delta $
and
$  \omega_H = E_\tau + \frac{1}{2}\Delta_B-\delta $,
where we have introduced
\beq
  \Sigma_B &=& (g_x^e + g_x^h) \mu_B B_x = E_B^e - E_B^h
  \nonumber\\
  \Delta_B  &=& (g_x^e - g_x^h) \mu_B B_x = E_B^e + E_B^h
  .
\eeq
Transforming to a rotating frame, we obtain the Hamiltonian
\begin{widetext}
\beq
  {\cal H} =
  \rb{
    \begin{array}{cccc}
      0 & 0 & \Omega_V^* e^{i\alpha} & \Omega_H^* e^{i \Delta_B t} \\
      0 & 0 & \Omega_H^*  & \Omega_V^* e^{-i \Sigma_B t+i\alpha} \\
      \Omega_V e^{-i\alpha} & \Omega_H  & \delta & 0 \\
      \Omega_H e^{-i \Delta_H t} & \Omega_V e^{i \Sigma_B t-i\alpha} & 0 & \delta
    \end{array}
  }
  \label{Ham}
  .
\eeq
\end{widetext}
The terms oscillating with frequencies $\Delta_B$ and $\Sigma_B$
describe the off-resonant driving of transitions $\mathbf{H}_2$ and
$\mathbf{V}_2$ respectively, both of which involve the state
$\ket{\tau_-}$. In deriving our analytic solution for the behaviour
of this system, it is these off-resonant transitions that we treat
approximately and it is therefore the magnitude of the quantities
$\Delta_B$ and $\Sigma_B$ that determine the accuracy of our
description.

\section{Adiabatic elimination of trion levels \label{SECae1}}

We now seek an compact, approximate description of the evolution of
the qubit sector and we shall proceed through the adiabatic
elimination of the trion levels.  This is appropriate for the situation we consider in which the detuning is sufficient that we only virtually occupy the trion level.
The equations of motion for the wave function coefficients under the
action of ${\cal H}$ are
\beq
  i \dot{c}_{x+} &=& \Omega_V^* e^{i \alpha} c_{\tau+}
    + \Omega_H^* e^{i\Delta_B t}c_{\tau-}
  \nonumber\\
  i \dot{c}_{x-} &=& \Omega_H^* c_{\tau+}
    + \Omega_V^* e^{-i\Sigma_B t+i\alpha}c_{\tau-}
  \nonumber\\
  i \dot{c}_{\tau+} &=& \delta c_{\tau+}
    + \Omega_V e^{-i\alpha}c_{x+} + \Omega_H c_{x-}
  \nonumber\\
  i \dot{c}_{\tau-} &=& \delta c_{\tau-}
    + \Omega_H e^{-i \Delta_B t}c_{x+}
    + \Omega_V e^{i \Sigma_B t - i \alpha} c_{x-}
    \label{eom}
    .
\eeq
The equations for $c_{\tau\pm}$ can be rewritten as
\beq
  \frac{\partial}{\partial t} \rb{c_{\tau+} e^{i \delta t}} &=&
  -i
  \left\{\frac{}{}
    \Omega_V e^{-i\alpha}c_{x+} + \Omega_H c_{x-}
  \frac{}{}\right\}
  e^{i \delta t}
  \nonumber\\
  \frac{\partial}{\partial t} \rb{c_{\tau-} e^{i \delta t}} &=&
  -i
  \left\{\frac{}{}
    \Omega_H e^{-i \Delta_B t}c_{x+}
  \right.
  \nonumber\\
    &&~~~~~~~
  \left.
    + \Omega_V e^{i \Sigma_B t - i \alpha} c_{x-}
  \frac{}{}\right\}
  e^{i \delta t}
  \label{ctdot}
  .
\eeq

We proceed by integrating by parts and assuming that the envelopes
$\Omega_{V,H}$ and thereby $c_{x\pm}$ are slowly-varying functions
of time.  We thereby arrive at an approximate expression for the
trion amplitudes in terms of the ground-state coefficients:
\beq
  c_{\tau +} &\approx &
    -
    \frac{\Omega_V e^{-i\alpha}}{\delta}
    c_{x+}
    -
    \frac{\Omega_H }{\delta}
    c_{x-}
    ,
  \nonumber\\
  c_{\tau -} &\approx &
    -
    \frac{\Omega_H e^{-i \Delta_B t}}{\delta-\Delta_B}
    c_{x+}
    -
    \frac{\Omega_V e^{i \Sigma_B t - i \alpha}}{\delta+\Sigma_B}
    c_{x-}
  \label{ct}
\eeq
The validity of these expressions are conditioned on the following
adiabatic constraints
\beq
  \left|\Omega_V c_{x+} \right|
  &\gg&
  \left|
    \frac{d}{dt} \rb{ \frac{\Omega_V c_{x+}}{\delta} }
  \right|,
  \nonumber\\
  \left|\Omega_H c_{x-} \right|
  &\gg&
  \left|
    \frac{d}{dt} \rb{ \frac{\Omega_H c_{x-}}{\delta} }
  \right|,
  \nonumber\\
    \left|\Omega_H c_{x+} \right|
  &\gg&
  \left|
    \frac{d}{dt} \rb{ \frac{\Omega_H c_{x+}}{\delta-\Delta_B}}
  \right|,
  \nonumber\\
  \left|\Omega_V c_{x-} \right|
  &\gg&
  \left|
    \frac{d}{dt} \rb{\frac{ \Omega_V c_{x-}}{\delta+\Sigma_B}}
  \right|,
  \label{const}
\eeq
which mean that, as well as the laser amplitudes being slowly
varying, the laser frequencies can not be too close to resonance
with any the transitions.

Assuming that these conditions are met, we can substitute
$c_{\tau\pm}$ from Eq.~(\ref{ct}) into the remaining equations of
motion~(\ref{eom}), and obtain a closed set of equations of motion
for the qubit coefficient $c_{x\pm}$:
\beq
  i \frac{d}{dt}
  \rb{
  \begin{array}{c}
    c_{x+} \\ c_{x-}
  \end{array}
  }
  = h(t)
  \rb{
  \begin{array}{c}
    c_{x+} \\ c_{x-}
  \end{array}
  }
\eeq
where $h(t)$ is the effective Hamiltonian acting solely on the qubit
space:
\beq
  h(t) =
  -
  \rb{
  \begin{array}{cc}
    \frac{|\Omega_V|^2}{\delta}
    + \frac{|\Omega_H|^2}{\delta-\Delta_B}
  &
    \frac{\Omega_V^*\Omega_H}{\delta}e^{i\alpha}
  \\
   \frac{\Omega_V\Omega_H^*}{\delta}e^{-i\alpha}
  &
    \frac{|\Omega_H|^2}{\delta}
    +\frac{|\Omega_V|^2}{\delta+\Sigma_B}
  \end{array}
  }
  \label{h(t)}
  .
\eeq
In writing this Hamiltonian, we have neglected further terms
oscillating with frequency $ 2 E_B^e$.  These terms are expected to
be negligible since, without them, the Rabi-frequency of the
effective model is $\Omega_V^* \Omega_H / \delta$, and thus provided
that
\beq
  2 |E_B^e| \gg |\Omega_V^*\Omega_H / \delta|
  \label{const2}
\eeq
these terms are rapidly oscillating in comparison and thus
approximately self-average to zero.
This effective two-level Hamiltonian $h(t)$ forms the basis of our
approach and provides a good account of the system as we will show.

%%%%%%%%%%%%%%%%%%%%%%%%%%%%%%%%%%%%%%%%%%%%%%%%%%%%%%%%%%%%%%%%%%%%%%%%%
\section{Time-evolution operator}

Given that the qubit evolves under the action of effective
Hamiltonian $h(t)$, we may once again make use  of the adiabaticity
of the system to obtain an approximate form for $U(t)$, the
time-evolution operator of the system.
Firstly, let us specify that the envelopes $\Omega_V(t)$ and
$\Omega_H(t)$ have the same shape but different amplitudes and write
$\Omega_H(t) = \nu \Omega_V(t)$, where $\nu$ is constant in time.
We can then write $h(t)$ in the form
\beq
  h(t) &=&
  \frac{1}{2}\lambda(t)  \bm{n}\cdot\bm{\sigma} + \mu
  \mathbbm{1}
\eeq
and can drop the constant term $\mu \mathbbm{1}$.  In general, both
the forefactor $\lambda$ and the unit vector $\bm{n}$ depend on
time, but since we are assuming the same shape for both pulses, the
axis ${\bf n}$ remains fixed, and all the time dependence is
contained in $\lambda(t)$. We can then approximate the time
evolution operator as
\beq
  U(t) =\exp
  \left\{-i /2 \Lambda(t) ~\bm{n}\cdot\bm{\sigma} \right\}
  \label{U(t)}
  ,
\eeq
where $\Lambda(t) = \int_{-\infty}^{t} dt' \lambda(t')$.  The final
output gate operator is $U=U(t\rightarrow \infty)$.

The operator $U$ represents an arbitrary single-qubit rotation.  The
rotation angle at time $t$ is given by
\beq
  \Lambda (t) &=&
  \frac{r}{\left|\delta(\Delta_B-\delta)(\delta + \Sigma_B)\right|}
  \int_{-\infty}^{t}\Omega^2(t') dt'
  ,
\eeq
where
\beq
  r^2 &=&
  \Sigma_B^2 (\Delta_B - \delta)^2 + \Delta_B^2(\delta + \Sigma_B)^2
  \nu^4
  \nonumber\\
  &&
  + 2(\Delta_B-\delta)(\delta+\Sigma_B)
  \nonumber\\
  &&~~~~
  \times
  \left[
    \Delta_B(\Sigma_B+2\delta)\nu^2 - 2 \delta(\delta + \Sigma_B)
  \right] \nu^2
\eeq
Note that this expression for $\Lambda (t)$ depends only on the ``area-under-the-pulse-squared'', and not on the details of the shape of the pulse.  This is typical of the adiabatic approach we are pursuing here.
For simplicity, we will consider a Gaussian pulse
envelope, $\Omega_V = A \exp\rb{-t^2/2T^2}$, in which case the total
angle is
\beq
  \Lambda &=&  \Lambda(t\rightarrow \infty)=
  \frac{ r A^2 \sqrt{\pi} T}
  {\delta(\Delta_B-\delta)(\delta + \Sigma_B)}
  .
\eeq
The axis of rotation has components
\beq
  n_1 &=& \cos\beta\cos \alpha
  \nonumber \\
  n_2 &=&- \cos \beta\sin \alpha
  \nonumber \\
  n_3 &=& \sin \beta
\eeq
where
\beq
  \tan \beta =
  \frac{
   \Sigma_B(\delta-\Delta_B) + \Delta_B(\delta + \Sigma_B)\nu^2
  }
  {
   2 \nu(\delta-\Delta_B)(\delta+\Sigma_B)
  }
\eeq
and $\alpha$ is the phase difference between the pulses.

Let us now describe two specific rotations as illustrative examples.
We consider first a rotation about the $\bm{3}$ axis, which
corresponds to a change in the relative phase of the qubit states.
Only a single laser pulse is necessary and thus we set $\nu=0$. The
rotation is then precisely about the $\bm{3}$ axis, $\bm{n} =
(0,0,1)$, by an angle of
\beq
  \Lambda &=& \frac{ \sqrt{\pi} A^2 \Sigma_B T}{\delta(\delta + \Sigma_B)}
  \label{Lnu0}
  .
\eeq

A rotation with the axis in the $\mathbf{1}$-$\mathbf{2}$ plane
includes a transfer of population between the two qubit levels.
This requires two lasers and, by setting the relative amplitude of
the two pulses according to
\beq
  \nu =
  \sqrt{
    \frac{\Sigma_B(\Delta_B-\delta)}
    {\Delta_B (\Sigma_B+\delta)}
  }
  ,
\eeq
we set $n_3=0$, and obtain a rotation axis exactly in the
$\bm{1}$-$\bm{2}$ plane. The direction of this axis is then
specified by the relative phase of the two lasers, $\alpha$.  In
order for $\nu$ to be real and finite we require that $\Delta_B \le
\delta < -\Sigma_B$ for $\Delta_B,\Sigma_B < 0$, as is the case
here.

%%%%%%%%%%%%%%%%%%%%%%%%%%%%%%%%%%%%%%%%%%%%%%%%%%%%%%%%%%%%%
\begin{figure}[t]
  \begin{center}
  \psfrag{F}{${\cal F}$}
  \psfrag{hDel}{$\delta$(meV)}
  \epsfig{file=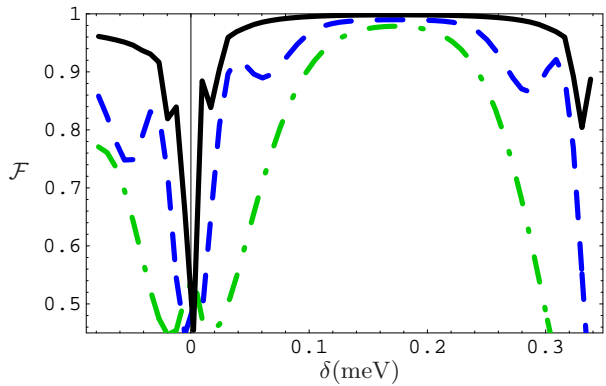, clip=true,width=0.95\linewidth}
  \caption{
    The fidelity ${\cal F}$, which reflects the degree to which our
    description approximates the exact evolution of the system, as a function
    of Raman detuning $\delta$.  Here we consider a
    $\pi$-rotation about the $\bm{3}$ axis, corresponding to a
    change in relative phase between the two qubit levels, using pulses
    of durations $T=5$~ps (green dash-dot), $15$~ps (blue dash)
    and $50$~ps (black solid line).  As is clear, for this type of rotation,
    excellent agreement (${\cal F}$ close to unity)
    can be obtained with even very short pulses ($T\approx5$~ps).
    The model parameters are
    $g_x^e = -0.46$ and $g_x^h = -0.29$ with a magnetic field of $8$~T.
    Laser parameters were $\nu=0$, with $A$ chosen to give $\Lambda=\pi$.
    \label{fidfig1}
 }
  \end{center}
\end{figure}
%%%%%%%%%%%%%%%%%%%%%%%%%%%%%%%%%%%%%%%%%%%%%%%%%%%%%%%%%%%%%%%%%%

%%%%%%%%%%%%%%%%%%%%%%%%%%%%%%%%%%%%%%%%%%%%%%%%%%%%%%%%%%%%%%
\section{Fidelity}

The preceding sections demonstrate that arbitrary single-qubit
rotations are possible within this set-up.  The important question
then arises as to how fast these operations can be performed.
This requires that we know the degree to which our approximate
description matches the actual behaviour of the system.
This we quantify with the fidelity between the calculated operation
and the result obtained by numerical integration of the Schr\"odinger
equation.
The fidelity is defined as \cite{poy97}
\beq
  {\cal F} =
  \overline{\ew{\Psi_\mathrm{in} |\widetilde{U}^\dag \rho_\mathrm{out}
  \widetilde{U}|\Psi_\mathrm{in}}}
  \label{fid}
  ,
\eeq
where the overline represents an average over all input states
$\ket{\Psi_\mathrm{in}}$, $\widetilde{U}$ is the predicted
operation, and $\rho_\mathrm{out}$ is the actual output density
matrix.  In evaluating the fidelity from our numerics, we have used the method described in Ref. \cite{pie02}.

We calculate the fidelities for the two $\Lambda=\pi$ rotations above: one
about the $\bm{3}$ axis and the other about the $\bm{1}$ axis.
In Fig.~\ref{fidfig1} we plot ${\cal F}$ as function of the laser
detuning $\delta$ for the rotation about the $\bm{3}$ axis for
several values of the pulse duration $T$.
This figure shows that, for the $\bm{3}$-axis rotation, even very
short pulses ($T\approx 5$~ps) can be used with fidelity close to
unity provided that the detuning $\delta$ is chosen appropriately.
%
%
%%%%%%%%%%%%%%%%%%%%%%%%%%%%%%%%%%%%%%%%%%%%%%%%%%%%%%%%%%%%%
\begin{figure}[t]
  \begin{center}
  \psfrag{F}{${\cal F}$}
  \psfrag{hDel}{$\delta$(meV)}
  \epsfig{file=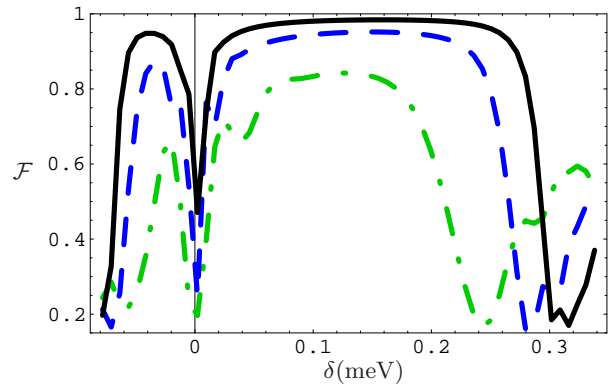, width=0.95\linewidth}
  \caption{  
    Same as Fig.~\ref{fidfig1} except that here the rotation is about the $\bm{1}$ axis
    and the pulse durations are: $T=20$~ps (green dash-dot), $50$~ps (blue dashes)
    and $100$~ps (black solid line).  Longer times are required to perform this
    operation, which involves a transfer of population between qubit
    levels.  Pulse parameters $\nu$ and $A$ were chosen to give the
    desired rotation axis and angle, $\Lambda=\pi$, for each value
    of the detuning $\delta$.
    \label{fidfig2}
 }
  \end{center}
\end{figure}
%%%%%%%%%%%%%%%%%%%%%%%%%%%%%%%%%%%%%%%%%%%%%%%%%%%%%%%%%%%%%%%%%%
%
In Fig.~\ref{fidfig2} we plot the same thing for the rotation about
the $\bm{1}$-axis and observe that longer pulse durations are
required to obtain high fidelities.
Irregularities in ${\cal F}$ occur in both these figures at values
of $\delta=0,\Delta_B,-\Sigma_B$.  These are the regions at which we
know from Eq.~(\ref{const}) that the perturbative theory breaks
down.

\begin{figure}[t]
  \begin{center}
  \psfrag{F}{${\cal F}$}
  \psfrag{tau}{$T$(ps)}
  \epsfig{file=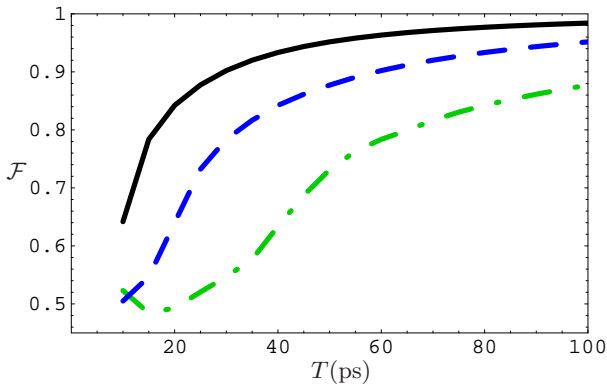, clip=true,width=0.95\linewidth}
  \caption{
    The maximum fidelity ${\cal F}$ for a $\Lambda=\pi$
    rotation about the $\bm{1}$-axis as a function of pulse duration $T$.  Results are
    shown for magnetic fields of $2$~T (green dash-dot), $4$~T (blue dash),
    and $B=8$~T (black solid line),
    with $g$-factors as in Fig.~\ref{fidfig1}.
    Fidelity increases  with both
    magnetic field and pulse duration.
    For B=$8$~T a fidelity of $95$\% can be obtained with pulse
    durations of $\approx 50$~ps.
    \label{fidvtaufig}
 }
  \end{center}
\end{figure}

We now examine the dependence on the applied magnetic field of the
maximum obtainable fidelity for a given pulse-duration, .  We
consider a $\Lambda=\pi$, $\bm{1}$-axis rotation, since this is the
more demanding rotation, and in  Fig.~\ref{fidvtaufig} plot the
results for several different magnetic fields.
As is clear, this field determines the pulse-duration required to
obtain a given fidelity.  For the parameters used here a field of
$8$~T gives a fidelity of 95\% for a pulse duration of about 50~ps.

The required pulse-duration is reduced if the $g$-factors are
larger.  In particular, we have used here an electronic
$g$-factor of $|g_x^e| = 0.46$.  This is quite small compared
with other measurements reported for InAs QDs, which have yielded
values of $|g_x^e| = 0.6$ \cite{ata06} and $|g_x^e|\approx 0.9$
\cite{hap02}. Use of such dots will significantly reduce either the
magnetic field or the pulse duration  required to perform
high-fidelity operations.  It should also be noted that a
$\pi$-rotation about the $\bm{1}$-axis is the worst case example and
that all other rotations can be achieved in less time.

In any case, these time scales are much shorter than the trion
lifetime in InAs dots which has been measured to be $1$~ns
or longer \cite{cor02,war05}. This, coupled with the fact that
the gate operation proceeds essentially adiabatically, means that
the effects of spontaneous emission from the trion has
negligible effect on the gate operation.
%%%%%%%%%%%%%%%%%%%%%%%%%%%%%%%%%%%%%%%%%%%%%%%%%%%%%%%%%%%%%
%%%%%%%%%%%%%%%%%%%%%%%%%%%%%%%%%%%%%%%%%%%%%%%%%%%%%%%%%%%%%%%%%%

%%%%%%%%%%%%%%%%%%%%%%%%%%%%%%%%%%%%%%%%%%%%%%%%%%%%%%%%%%%%%%%%%%
\section{Hole mixing}

We now include the effects of hole-mixing in our analysis.
The dominant mixing terms in the Luttinger Hamiltonian
\cite{lut55,bro85} mean that, rather than the `bare' heavy-hole
states $\ket{\sfrac{3}{2},\pm\sfrac{3}{2}}$, the actual states are
better approximated as
\beq
  \ket{H^{\pm}_z} &=&
  \cos \theta_m \ket{\sfrac{3}{2},\pm\sfrac{3}{2}}
  - \sin \theta_m e^{\mp i\phi_m}\ket{\sfrac{3}{2},\mp\sfrac{1}{2}}
  \nonumber\\
  \label{Hpm}
\eeq
where $\ket{\sfrac{3}{2},\mp\sfrac{1}{2}}$ are light-hole states,
$\theta_m $ and $\phi_m$ are mixing angles, and we have used the growth
direction as the quantization axis.

In terms of orbital and spin degrees of freedom, the valence-band
electron states are \cite{bro85,bronote}
\beq
  \ket{\sfrac{3}{2},\pm\sfrac{3}{2}}_e &=& \ket{\pm 1,\pm\sfrac{1}{2}}_e
  \nonumber\\
  \ket{\sfrac{3}{2},\pm\sfrac{1}{2}}_e &=& \sqrt{\sfrac{1}{3}}\ket{\pm1,\mp\sfrac{1}{2}}_e
    \pm \sqrt{\sfrac{2}{3}}\ket{0,\pm\sfrac{1}{2}}_e
    \nonumber\\
\eeq
where on the right-hand side, the first index corresponds to the
orbital degree of freedom,
\beq
  \ket{\pm 1} = \sqrt{1/2} \rb{\ket{X}\pm i\ket{Y}}
  ;\quad
  \ket{0} =\ket{Z}
  ,
\eeq
and the second to the electron spin $\pm1/2$.  The make-up of these
states means that, for light propagating in $z$-direction,
transitions involving the light-hole components of $\ket{H_z^\pm}$
are possible and that they acquire an extra factor of $\sqrt{1/3}$
in the matrix element.

We are now in a position to consider the Hamiltonian of the system
including hole-mixing, for which we use the same basis as for
Eq.~(\ref{Ham}), except that now we use the hole-states
$\ket{H_x^\pm} = 1/\sqrt{2}\rb{\ket{H_z^-} \pm \ket{H_z^+}}$.
If we illuminate the dot with the same linear polarizations as
before, hole-mixing means that all four transitions are driven by
each polarization and the Hamiltonian of the system is far more
complicated than that considered in the foregoing analysis. However,
by adjusting the polarizations of the two lasers, we can reach a
situation where the Hamiltonian is identical to that of
Eq.~(\ref{Ham}), but with renormalized parameters.  The single qubit
rotations then follow directly.

The two laser polarizations that accomplish this are
  $\mathbf{V}' = 2^{-1/2}(\svec_+ + e^{i\mu_+}\svec_-)$
and
and
  $\mathbf{H}' = 2^{-1/2}(\svec_+ - e^{-i\mu_-}\svec_-)$
with the phases
\beq
  e^{i \mu_\pm}  =
  \frac
    {\sqrt{3}\cos \theta_m \pm \sin\theta_m e^{\pm i\phi_m}}
    {\sqrt{3}\cos \theta_m \pm \sin\theta_m e^{\mp i\phi_m}}
  .
\eeq
Note that these two polarization are non-orthogonal in general.
Using these polarizations we obtain a Hamiltonian the same form as
Eq.~(\ref{Ham}) but with $\Omega_V$ and $\Omega_H$ being replaced
by
\beq
  \widetilde{\Omega}_{V'} &=& \Omega_{V'}
    \frac{1+ 2 \cos 2\theta_m}{3\cos \theta_m + \sqrt{3}e^{-i\phi_m}\sin\theta_m}
    ,
  \nonumber\\
  \widetilde{\Omega}_{H'} &=& \Omega_{H'}
    \frac{1+ 2 \cos 2\theta_m}{3\cos \theta_m - \sqrt{3}e^{i\phi_m}\sin\theta_m}
  ,
\eeq
where $\Omega_{V'}$ and $\Omega_{H'}$ are the actual Rabi
frequencies of the two transitions.
Use of these new polarizations therefore allows us to
circumvent the effects of hole mixing and proceed directly as
outlined in the previous sections.

%%%%%%%%%%%%%%%%%%%%%%%%%%%%%%%%%%%%%%%%%%%%%%%%%%%%%%%%%%%%%%%%%%
\section{GaAs dots}

We now describe briefly the application of the above theory to QDs
in which the trion level does not split under in-plane magnetic
field.  This was the situation studied in Ref.~\cite{che04}, but the
approach we pursue here is different and leads to
an improved description of this system.

With the in-plane hole $g$-factor equal to zero, and with no
hole-mixing, illumination with $\svec_+$-circularly-polarized light propagating in the
growth direction excites a trion state $\ket{\tau_z+}$ containing a
spin-up heavy hole $\ket{\Uparrow}$ aligned in the growth direction.
We employ two such $\svec_+$ lasers set to a
detuning $\delta$ from the Raman transition between the two electron
ground states $\ket{x\pm}$.  A full description is given in
Ref.~\cite{che04}, where it is shown that in the basis $\left\{
\ket{x+}, \ket{x-},\ket{\tau_z+}, \right\}$, the Hamiltonian for the
system in the rotating frame is given by
\begin{widetext}
\beq
  {\cal H} =
  \rb{
    \begin{array}{ccc}
      0& 0 & \Omega^*_\uparrow e^{i\alpha} + \Omega^*_\downarrow e^{i E_B^e t}\\
      0 &0 & \Omega^*_\uparrow e^{- iE_B^e t + i\alpha} +\Omega^*_\downarrow  \\
      \Omega_\uparrow e^{-i\alpha} + \Omega_\downarrow e^{-i E_B^e t}
        & \Omega_\uparrow e^{i E_B^e t - i\alpha} + \Omega_\downarrow
          & \delta
     \end{array}
  },
\eeq
\end{widetext}
where $ \Omega_\uparrow(t)$ and $ \Omega_\downarrow(t)$ are the Rabi
frequencies of the two $\svec_+$-induced transitions, and $E_B^e$ is the Zeeman
splitting of the electron.

From this Hamiltonian we can derive an effective two-level model for
the qubit sector exactly as in the previous sections.  We obtain
\beq
  h_1(t) =
  -\rb{
  \begin{array}{cc}
  \frac{|\Omega_\uparrow|^2}{\delta}
  +
  \frac{|\Omega_\downarrow|^2}{\delta - E_B^e}
  &
  \frac{\Omega^*_\uparrow \Omega_\downarrow e^{i \alpha}}{\delta}
  \\
  \frac{\Omega_\uparrow \Omega_\downarrow^* e^{-i \alpha}}{\delta}
  &
  \frac{|\Omega_\uparrow|^2}{\delta + E_B^e}
  +
  \frac{|\Omega_\downarrow|^2}{\delta}
  \end{array}
  }
  .
\eeq
This has the same form as Eq.~(\ref{h(t)}) with the substitutions
$\Omega_V \rightarrow \Omega_\uparrow$, $\Omega_H \rightarrow
\Omega_\downarrow$ and $\Sigma_B,\Delta_B\rightarrow E_B^e$.  The
axis and angle of the corresponding single-qubit rotations follow
directly.

Although derived in a different way, the work of Chen {\it
et al.} \cite{che04} essentially posits
that the qubit is governed by the effective Hamiltonian
\beq
  h_2(t) =
  -
  \frac{1}{\delta}
  \rb{
  \begin{array}{cc}
  {|\Omega_\uparrow|^2}
  &
  {\Omega^*_\uparrow \Omega_\downarrow e^{i \alpha}}
  \\
  {\Omega_\uparrow \Omega^*_\downarrow e^{-i \alpha}}
  &
  {|\Omega_\downarrow|^2}
  \end{array}
  }.
\eeq
This differs from the Hamiltonian $h_1$ by the absence of the
on-diagonal terms with $\delta \pm E_B^e$ in the denominator that describe the
off-resonant effects of the qubit levels being driven by both
lasers.
These terms have a significant impact on the accuracy of the
description with pulses as short as we use here. With parameters for
which the description based on $h_1(t)$ has a fidelity of 95\%, that
based on $h_2(t)$ will typically have a fidelity of the order of
75\%.  This shows that the treatment of these off-resonant terms is
essential to the proper understanding of single-qubit rotation in
QDs and that the adiabatic elimination approach pursued here
provides just such a treatment.

%%%%%%%%%%%%%%%%%%%%%%%%%%%%%%%%%%%%%%%%%%%%%%%%%%%%%%%%%%%%%%%%%%
\section{Summary}

We have presented a theory of single-qubit rotations in InAs quantum
dots.  This theory provides a description of the behaviour of the
system to high fidelity for realistic dot parameters and short laser
pulses.

Our theory shows how two Raman-detuned lasers can be used to obtain
arbitrary single-qubit rotations, and that this can be  accomplished
with laser pulses the duration lasting a few tens
of picoseconds.
We have also shown that hole-mixing can be simply incorporated into
this scheme through a change in laser polarizations.

This work differs from that presented in  Ref.~\cite{che04} not only
in that it applies to systems in which the hole has an in-plane
g-factor but also in the methodology.  The adiabatic elimination
approach that we have utilized here provides better a account of the
system because it includes the effects of the
unwanted off-resonant transitions.

\medskip
This work was supported by ARO/NSA-LPS.

%%%%%%%%%%%%%%%%%%%%%%%%%%%%%%%%%%%%%%%%%%%%%%%%%%%%%%%%%%%%%%%%%%%%%%%

%%%%%%%%%%%%%%%%%%%%%%%%%%%%%%%%%%%%%%%%%%%%%%%%%%%%%%%%%%%%%%%%%%%%%%%

\begin{thebibliography}{99}
\bibitem{bronote}  We have used differnt global phases for some of the wave
    functions as compared with Ref.~\cite{bro85}.
\end{thebibliography}
\end{document}